\documentclass[aps,prl,twocolumn]{revtex4}
\usepackage{graphics}
\def\lsp{{\it LSP }}
\def\lsps{{\it LSP}s }

\begin{document}

\pagestyle{empty}


\title{Assessing the Impact of Student Learning Style Preferences}

\author{Stacey M. Davis}
\author{Scott~V.~Franklin}
\email[]{svfsps@rit.edu}
\homepage[]{http://piggy.rit.edu/franklin/}
\affiliation{Dept. of Physics, Rochester Institute of Technology}

\date{Aug, 2003}

\begin{abstract}

Students express a wide range of preferences for learning
environments.  We are trying to measure the manifestation of learning
styles in various learning environments.  In particular, we are
interested in performance in an environment that disagrees with the
expressed learning style preference, paying close attention to social
(group vs. individual) and auditory (those who prefer to learn by
listening) environments.  These are particularly relevant to
activity-based curricula which typically emphasize group-work and
de-emphasize lectures.  Our methods include multiple-choice
assessments, individual student interviews, and a study in which we
attempt to isolate the learning environment.
\end{abstract}

%\keywords{}

\maketitle
\begin{center}{\bf Introduction}\end{center}

\vskip -0.1in A learning style is a biologically and developmentally imposed set of
personal characteristics that make the same teaching (and learning)
methods more effective for some and less effective for others
\cite{Dunn89}.  These include techniques, approaches, and
processes\cite{Dunn90}, but also innate physiological factors,
experience, habit, and training.  Learning styles are consistent with
personality types, but there is more to one's learning style than
personality.  Common learning styles diagnostics range from the
Jungian-based Myers-Briggs personality type test\cite{MyersBriggs} to
more detailed attempts to discern environmental and physiological
effects \cite{Dunn2, Felder}.  As these rely on student
self-reporting, they suffer greatly from the fact that students often
don't know when they learn, let alone how they best learn.  Students
that claim to learn best by listening often mean that they are most
comfortable following a competent lecturer.  This comfort often does
not correlate with learning (in fact, it sometimes is anti-correlated
with learning)\cite{Learning}.  As such, it is perhaps more accurate
to talk about learning style {\it preferences (LSP)}.

To accommodate different \lsps, many research-based physics curricula
\cite{WS, ILDS, Tutorials} present information in a variety of
representations \cite{Larkin02,Larkin01}.  Motion, for example, is
described with words, pictures, graphs, and, ultimately, equations.
As measured by standard conceptual evaluations \cite{Thornton98},
these courses produce learning gains significantly larger than
traditional courses.  These learning gains are experienced by all
segments of the class, with stronger students benefiting the most by
the reformed curricula \cite{Beichner}.

A study on deaf students \cite{Lang99} found a correlation between
learning style preference and course grade, with students who have a
more participatory approach to learning earning higher grades.  Dunn,
et al. \cite{Dunn90} also found that accommodating learning styles
could boost student performance by almost one standard deviation.
Felder has analyzed \cite{Felder88,Felder02} student performance in
introductory engineering classes in the context of {\it LSP}s.  He found
that extroverts performed almost one full letter grade higher than
introverts, and speculated that the cooperative learning benefited the
extroverts.  He also found a significant gender gap \cite{Felder02} in
performance between students who tend to make judgments subjectively
and personally (Jungian {\it feelers}), but no gap between those who
approach learning more objectively (Jungian {\it thinkers}).
Addressing \lsps may begin to remedy the under-performance of women in
introductory physics classes \cite{McCulloughProc,McCulloughProc2}.

\begin{center}{\bf Multiple-choice \lsp assessments}\end{center}

\vskip -0.07in Dunn and Dunn have developed the {\it Productivity
Environmental Preference Survey (PEPS)} \cite{Dunn2} which
incorporates environmental, perceptual, and sociological preferences.
The {\it PEPS} test, a 100-item, 5-point Likert scale, evaluation,
breaks from the traditional either/or classification of type, instead
reporting a level of compatibility with a particular style.
Compatibility with seemingly contradictory styles is possible.  For
example, an individual may have a high compatibility with a group
learning environment as well as an individual environment.  Relevant
perceptual elements include auditory, tactile and verbal kinesthetic,
and visual picture.  Preferences for group or individual, tactile or
verbal envirnoments might have important ramifications in a
group-based introductory physics course.

Rundle's {\it Building Excellence (BE)} exam \cite{BE} is similar to
the {\it PEPS} test.  It is an 111-item questionnaire that uses a
5-point Likert scale.  It expands the social dimension to include
small teams of 2-3 people, as opposed to just individual or group
preferences.  In addition, it can be administered online.

\begin{center}{\bf Correlating Course Grade with {\it LSP}}\end{center}

\vskip -0.05in The {\it Building Excellence} exam was administered to 390 students
enrolled in the first quarter of RIT's three-quarter calculus-based
introductory physics course.  98 students participated in the fall of
2002 and 292 participated in the winter of 2002-2003.  The test was
administered on-line, so students could take it at their convenience
and it did not detract from class time, although all students that
took the test did so within the first 2 weeks of class.  We have not
investigated whether the classroom activities can influence student
response on Learning Style assessments; such a study would be quite
interesting.  A breakdown of student performance is shown in Table
\ref{grade}.  The average class grade was the same in the fall
quarter, but students in the traditional sections in the winter had a
higher average grade (2.84 to 2.43).  Our current analysis looks for
differences between students in similar environments, so this
difference is not a problem.  In order to compare performance between
students in different environments we compare the deviation from mean
section grade.  This seems to remove the artifact caused by the
different average grade of different sections.

\begin{table}
\begin{tabular}[c]{||c||c||c||}
\hline\hline
&Fall 2002 & Winter 2002-3 \\
\hline\hline
\begin{tabular}{c}
	\parbox{0.4in}{\begin{center}\vskip 0.2in N \\ 
$<G>$\end{center}}
\end{tabular}
&
\begin{tabular}{c|c}
SCALE-UP & Lecture \\
\hline
55 & 43 \\
3.44 & 3.48
\end{tabular} & 

\begin{tabular}{c|c}
SCALE-UP & Lecture \\
\hline
41 & 251\\
2.43 & 2.84 
\end{tabular} \\

\hline\hline
\end{tabular}
\caption{\label{grade}Average course grades $<G>$ for students who
 took the {\it Building Excellence} survey.}
\end{table}

\begin{center}{\bf Social Environment}\end{center}

\vskip -0.05in The {\it BE} test gauges compatibility with three
different sociological styles, alone/pairs, small groups (3-4
students), or in teams (4 or more).  Table \ref{group}, combining
students from the fall and winter quarters, shows that there was
little difference in final class grade in either SCALE-UP or
traditional sections.  We hypothesize that students mold their
environment to match their preferences.  Students in traditional
classes who prefer group interactions might satisfy this need by
formin study groups.  Similarly, students in SCALE-UP classes who
prefer individual learning might find a niche within their group.

\begin{table}
\begin{tabular}[c]{||c||c||c||}
\hline\hline
&SCALE-UP & Traditional \\
\hline\hline
\begin{tabular}{c}
	\parbox{0.4in}{\begin{center}\vskip 0.2in \% \\ 
$<G>$\end{center}}
\end{tabular}
&
\begin{tabular}{c|c|c}
Alone & Group & Team \\
\hline
80 & 53 & 51 \\
2.37 & 2.37 & 2.30
\end{tabular} & 

\begin{tabular}{c|c|c}
Alone & Group & Team \\
\hline
77 & 55 & 55\\
2.92 & 2.97 & 2.91
\end{tabular} \\

\hline\hline
\end{tabular}
\caption{\label{group} Average grade $<G>$ for students expressing
compatibility with individual, group, or team environments.  No
correlation between performance and preference is seen.  Students with
a strong preference for individual environments do not fare worse in
the SCALE-UP environment, where group work is common.}
\end{table}

\begin{center}{\bf Auditory Learning}\end{center}

\vskip -0.05in Of particular interest to many faculty are auditory learners, or those
who claim to learn best by listening.  Unlike the social dimension,
the auditory dimension is exclusive; learners have either high,
neutral, or low aptitudes for auditory environments.  We looked for a
depressed average grade in high-auditory learners in SCALE-UP classes
and the converse in traditional classes.  As table \ref{aud} shows,
however, there is no apparent correlation between auditory preference
and grade.  There may be some self-selection here, as those with a
preference for auditory environments may choose traditional sections
over SCALE-UP sections.  The data, however, show little benefit from
this choice.

\begin{table}
\begin{tabular}[c]{||c||c||c||}
\hline\hline
&SCALE-UP & Traditional \\
\hline\hline
\begin{tabular}{c}
	\parbox{0.4in}{\begin{center}\vskip 0.2in \% \\ 
$<G>$\end{center}}
\end{tabular}
&
\begin{tabular}{c|c|c}
Auditory & Neut. & Low \\
\hline
28 & 40 & 31 \\
2.44 & 2.35 & 2.41
\end{tabular} & 

\begin{tabular}{c|c|c}
Auditory & Neut. & Low \\
\hline
38 & 38 & 23\\
2.99 & 3.03 & 2.69
\end{tabular} \\

\hline\hline
\end{tabular}

\caption{\label{aud} Average grade for students expressing a strong,
neutral, and low preference for auditory learning.  The
under-performance of low-auditory learners in traditional settings is
not statistically significant ($p=0.1$).}
\end{table}

Little correlation was found between course grades and any preference
as expressed on the {\it Building Excellence} exam.  There are several
possible explanations for this.  The final course grade may be too
coarse a measurement of learning to distinguish this effect.  Student
preferences may not, in fact, align with the environment that best
produces learning (consistent with \cite{Learning}).  Finally,
students may find ways to apply their particular learning styles
regardless of course structure.

\begin{center}{\bf Student Interviews}\end{center}

\vskip -0.05in The ability of students with strongly expressed preferences against
group learning appeared to succeed in the seemingly discordant
SCALE-UP environment.  One student, in particular, had an interesting
combination of \lsps and agreed to be interviewed several times
throughout the quarter.  ``Max's'' {\it BE} scores indicated a low
compatibility for learning in small groups, an aversion to auditory
learning, and a strong dislike for for authority-driven methods.  In
class, Max's ostensible participation was very limited, and frequently
his partners would turn and talk amongst themselves, leaving Max on
the periphery.  At the same time, his perceptual \lsp dimensions
classified Max as one who is internal and tactile kinesthetic, meaning
he learns by verbalizing to himself or to others and needs to be
actively doing something.  This tactile kinesthetic need may or may
not be specific to the task, and Max was often seen doodling, which
may have satisfied this need.

Max strongly preferred the SCALE-UP classroom to the traditional one
(he had dropped out of a previous traditional class), saying

\begin{center}\parbox{2.75in}{\small I learn a lot better with hands-on and
group activities.  As we got into the class, I realized that I
understood things a lot better, and I didn't know why. I kind of paid
more attention to it and I realized that we were explaining stuff to
each other and teaching each other.}
\end{center}

Max rarely spoke out in class, but saw himself participating in his
group although, as noted, his group did not share this view.  Max
included himself when describing group activities with statements like
``Here's where we are measuring the force...'', ``We're all
interacting, doing the same thing...'', or ``We're solving
problems...''.

Max maintained an above average grade (B) throughout the quarter,
falling at the end to a high C.  His {\it FMCE} post-test score of 60\%
was at the class average (Max did not take the pre-test so no
normalized gain can be calculated).  Especially when compared with his
experience with lecture-based course (he withdrew), Max's story in
SCALE-UP can be considered a success despite the extreme mis-match
between expressed preference and environment.

\begin{center}{\bf Isolating Learning Style Dimensions}\end{center}
\vskip -0.05in As many research-based curricula \cite{WS,Tutorials,ILDS} have
reported significant learning gains, often attributed in part to the
group work, the question of learning styles vis a vis group
interactions is important.  Specifically, are there students who learn
best individually and, if so, how do they fare in group activities?  A
related question involves the stronger students.  A common fear
amongst faculty skeptical of group work is that the stronger students
in a group will carry along their less capable partners.  Work by
Beichner \cite{Beichner} and others has shown that in fact stronger
students benefit most from the new activities, and a plausible
explanation is that the process of explaining ideas to partners
actually helps learning (along the idea that one doesn't learn until
one teaches).  The proof, however, is rather indirect.  It is not
clear whether the student learning is improved because of the group
activities or from the research-based activities all students are
asked to perform.

{\it Methodology}

Student volunteers were solicited and paid to spend two hours working
through activities and taking various \lsp assessments.  Students were
required either to have taken introductory calculus-based physics in
the previous 2 years or to be currently enrolled in the course.  After
a short pre-test, students spent approximately 40 minutes on each of
two activities.  A post-test concluded the session.  In the first
hour, half of the students worked on a worksheet in groups of three
while the other half worked on the same worksheet alone.  In the
second hour, the groups switched. To reduce the chance that students
would be familiar with the topics, we chose activities involving
buoyancy, a topic typically outside the typical introductory physics
curriculum.  Related activities included hydrostatics, which research
has shown students to struggle with.  Activities had been developed as
part of the {\it Explorations in Physics} \cite{EiP} curriculum and
were adapted for this research.

The pre-test incorporated those questions from the {\it Building
Excellence} survey which probed the social dimension, and assessed
student preferences for group or individual activities.  Students were
also given the Keirsey Temperament Sorter, a 70-item questionnaire, to
assess personality types.  Pre- and post-content tests were devised
and tested on 1st-year physics majors who were not participating in
the study.  This test confirmed that the topics chosen were at the
appropriate level but also commonly misunderstood.

Students were randomly divided into groups of 12.  Of these, 6
students worked on an activity alone, and 6 were split into groups of
3.  After working on an activity for 45 minutes, the groups of 12
switched lab rooms.  Those that first worked individually now worked
in a small group, and those who first worked in a small group, now
worked individually.  The activity guide contained 2-3 self-contained
experiments that students could perform with little prior preparation.
Students were asked to make predictions, record data, posit
explanations and imagine applications for the ideas they develop.
Students were asked to record complete answers whether they worked
alone or in a group.  When they had spent 45 minutes on each of the
two topics, students were then given a post-test.

{\it Preliminary Results}
As with the previous study involving course grades, little correlation
between personality types, sociological learning style preference and
performance on pre- and post-tests were found.  We offer some
possible explanations for this lack of correlation, recognizing that
there may be many more.  Possible explanations include,
\begin{itemize}
\item \vskip -0.05in the expressed learning style preference may bear little
connection with the environment in which the student best learns
\item \vskip -0.1in college students may effectively activate other learning
resources when placed in a less preferred environment
\item \vskip -0.1in activities might need to be refined to fit within the alloted
forty-five minutes, or the chosen topics may be inappropriate
\item \vskip -0.1in pre- and post-tests are too coarse to measure improvement in
student understanding
\item \vskip -0.1in  8am on a Saturday morning may be too early to start any study
involving college students

\end{itemize}

\begin{center}{\bf Summary}\end{center}
\vskip -0.05in Learning and the educational setting is a very complicated balance of
learning styles, teaching styles, personality types, environmental
factors, innate physiological and psychological factors, motivation,
socioeconomic backgrounds, culture, and numerous other factors that
may effect the learner.  While common assessments that have been
validated for internal consistency do produce some discrimination
between different students, there appears to be little significant
correlation between learning style preference and performance (as
measured by course grade) in different learning environments.  This is
greatly complicated by the fact that classes, extending over a
ten-week quarter, expose students to many different environments.  In
addition, students possibly seek out-of-class environments that more
closely match their preference.  (This will be the subject of an
upcoming study in which we will ask students about their out-of-class
activities and look for correlations with their expressed {\it LSP}.)
Attempts to isolate students in a restricted environment do not yet
produce discrimination in learning, although we believe this
methodology, with significant refinement, shows promise.  Finally, by
studying individual students with extreme preferences we may gain
insight into the manner in which different students learn.  Our crude
analysis seems to indicate that we are not harming students by placing
them in the educational setting that might not best suit their
aptitudes.

\begin{acknowledgments}
SVF is grateful to Teresa Larkin for bringing attention to Learning
Style Preference research and assessment tools as well as useful
discussions.  This work has been supported by the National Science
Foundation under Grant No. 0116795.

\end{acknowledgments}
\vskip -0.2in \bibliography{/home/franklin/Writings/Reference_lists/references}
\end{document}